# Probing spin dynamics of 2D excitons with twisted light

Aswini Kumar Pattanayak[1†], Pritam Das[1†], Devarshi Chakrabarty[1], Avijit Dhara[1], Shreya Paul[1], Satyait Maji[1], Maruthi Manoj Brundavanam[1], Sajal Dhara[1*]

[1]*Department of Physics, Indian Institute of Technology, Kharagpur, 721302, India*

\* sajaldhara@phy.iitkgp.ac.in

[†] *A.K.P. and P.D. contributed equally to this paper*

**Abstract:** We propose a mechanism of intravalley spin-flip scattering in spin-valley coupled two-dimensional systems by transferring momentum of light into exciton center of mass using optical vortex (OV) beams. By varying the dispersion of light using the topological charge of OV beam, we demonstrate a unique approach to control the intra-valley spin-flip scattering rate of excitons. From our photoluminescence measurements, we demonstrate that the intravalley scattering rate in W-based TMDs can be tuned externally by OV beams. Variation of photoluminescence intensity with topological charges shows a crossover temperature (>150K), indicating competitions among time scales involving radiative recombination, spin-flip scattering, and thermal relaxations. Our proposed technique utilizing a structured light beam can open up a new approach to explore the physics of excitons in 2D systems.

**Introduction:**

Monolayer transition metal dichalcogenides (TMDs) with broken triangular sublattice symmetry originate intriguing valley physics, where electron spin is coupled with the valley degree of freedom due to high spin-orbit splitting of valence and conduction bands[1–5]. While the time-reversal symmetry ensures the valley degeneracy, the total spin momentum of strongly bound excitons determines the energy splitting between the bright ($X^b$) and dark ($X^d$) exciton states[6–11].

Dark excitons have been probed earlier using in-plane magnetic filed[8] and via out-of-plane polarized light excitation and detection[9,12]. However, no such technique has been implemented so far to tune the intravalley spin-flip scattering [13,14] externally between $X^b$ and $X^d$ states during steady-state near-resonant excitation.

In this work, we utilize the fact that in-plane momentum or dispersion of light can be varied with the topological charge ($l$) of optical vortex (OV) beam [15,16] and demonstrate a unique approach to excite excitons at higher momentum states which can be controlled externally. A unique attribute of OV beams is the orbital angular momentum (OAM) $l\hbar$ per photon [17,18] which finds applications across different branches of science[19–21]; among them is the transfer of OAM from photon to atomic electrons[22,23]. More recently OV beams have been utilized to probe the anomalous dispersion of excitons in monolayer $MoS_2$[24] via a blue shift of exciton emission energy with increasing $l$. The Radiative recombination of $X^b$ which shows up as the PL intensity is governed by the population ratio between $X^b$ and $X^d$ states observed in steady state measurements which is limited by intravalley spin flip scattering at fixed temperature (T)[13,14]. Intravalley spin-flip scattering of the bound electron allows transitions between the $X^b$ and $X^d$ states; therefore, it can be probed by steady-state PL intensity variation at fixed T with an external parameter affecting the intravalley spin-flip scattering. By keeping T fixed for a given sample, the phonon and impurity contributions[25,26] can be eliminated from the variation of the optical response with respect to the external parameter. The intravalley scattering rates depend quadratically on excitons center of mass momentum,[27–29] acquired from the small spread of in-plane photon momentum within the light cone under the resonant excitation.

**Results and Discussion**

A schematic of the sample, illuminated by an OV beam generated by a spatial light modulator and focused by an objective lens of numerical aperture 0.7, is shown in Figure 1a. An optical image of WS$_2$ monolayer excited with OV beam of $l = 2$ is shown as the inset. With increasing $l$ of the OV beam, the in-plane momentum of photons increases, facilitating the transfer of momentum to excitons, as shown in the schematic (Figure 1b) for $l = 0$ and 2. Excitons generated at higher momentum undergo enhanced intravalley scattering involving spin-degree of freedom between bright and dark excitons. A detailed analysis has been presented in the subsequent section. The total incident power on the sample was kept constant at ~500 nW to ensure no sample heating. We choose W-based TMDs, WS$_2$, and WSe$_2$, where the $X^d$ states lie below $X^b$ (considering only the charge-neutral A exciton in this work) resulting in PL quenching with decreasing T. At 150 K, where the PL intensity is finite for both the materials, we observe a similar reduction of intensity due to $X^b \to X^d$ transitions via intravalley spin-flip scattering with charge variation of OV beam. As shown in Figure 1c, the intensity of the WSe$_2$ exciton peak at 1.735 eV decreases with increasing $l$, and a similar trend is observed for WS$_2$ exciton peak (2.07 eV) as shown in Figure 1d. 520 nm and 660 nm excitation CW laser source has been used to excite WS$_2$ and WSe$_2$ monolayers respectively. Such intensity variation has been observed throughout the low-T regime for both WS$_2$, as shown in the representative data at 50 K in Supporting Information Figure S2. Intriguingly, a reversal of this intensity variation is observed at higher T, as shown in Figure 1e for WS$_2$. To rule out the effect of intensity variation arising due to spatial inhomogeneity of the sample, we have performed the experiment in a controlled way in two other WS$_2$ samples. We observed the PL intensity variation due to $l$ to be well dominant over spatial inhomogeneity as described in Figure S(3-5). The excitation and collection efficiency of the OV beam for different OAM is identical as shown in Figure S7. For all PL intensity measurements, the spectrometer resolution is kept constant at 3 meV unless it is mentioned. The reproducibility of these observations has been verified with another three samples of WSe$_2$, as shown in Supporting Information Figure S8. We attribute this intensity variation with charge of OV beam to the variation of intravalley spin-flip scattering rate and the variation of electric field distribution with $l$, which has been discussed below.

The initial momentum of excitons is determined by the in-plane momentum component ($k_\parallel$) of incident photons. The distribution of in-plane photon momentum and electric field components are analyzed at the focal plane of a high numerical aperture objective lens by angular spectrum representation[30] starting from order zero (Gaussian) to higher-order OV beams, which we assume to be a Laguerre-Gaussian beam for our calculations. It is observed that the most probable momentum of photon distribution shifts towards the higher momentum side with increasing $l$, as shown in Figure 2a. In-plane momentum distribution of excitons follows the momentum distribution of photons of OV beams to conserve momentum during excitation. Hence, root mean square momentum ($q_{rms}$) of exciton as a function of $l$, is estimated and considered as $q$, as shown in the inset of Figure 2a. Our simulation results reveal that excitons centre of mass momentum ($q$) increases with the increasing charge of the OV beam - enabling the transfer of the momentum of light into the excitons. Spin-flip scattering due to phonon and impurity scattering via Elliott-Yafet[31,32], Dyakonov-Perel[32] mechanisms contribute significantly, however, it remains constant at a fixed temperature. The spin-flip scattering by an effective magnetic field originated from short-range electron-hole exchange interaction (EHEI) by the de Andrada e silva and La Rocca mechanism is responsible for the momentum dependent intravalley scattering rate, written as, $\gamma^{intra} \propto (q^2)$ [33]. A brief theoretical description on momentum dependent intavalley scattering rate has been discussed in the Supporting Information. The dipole orientations of bright and dark excitons are parallel and perpendicular to the monolayer plane respectively[12,34,35]. Thus, exciton generation rates ($g_b$, $g_d$) are proportional to the intensity corresponding to in-plane ($|E_\rho|^2$) and out-of-plane ($|E_z|^2$) electric fields, which can vary with $l$ in this experiment. The field distribution of $E_\rho$ and $E_z$ is simulated in the focal plane for different $l$ [Figure 2b]. To quantify the variation of exciton generation rate for OV beam ($l \neq 0$) with respect to gaussian beam ($l = 0$), we calculate real space integration of $|E_\rho(l)|^2, |E_z(l)|^2$ and plot the relative change in generation rate $\left(\Delta g_d(l) = \frac{g_d(l) - g_d(0)}{g_d(0)}\right)$ and $\left(\Delta g_b(l) = \frac{g_b(l) - g_b(0)}{g_b(0)}\right)$ as shown in Figure 2c. It is observed that the enhancement in $\Delta g_d(l)$ is around 100% whereas, the reduction in $\Delta g_b(l)$ is around 10% for a change of $l$ from 0 to 5. So using higher order OV beams, more dark excitons and less bright excitons are excited directly. The variation of intravalley spin-flip scattering and generation rates play a crucial role to explain the T-dependent intensity variation with $l$.

To understand the crossover regime, a T-dependent PL measurement is done at the different charges of OV beam as shown in Figure 3a ($WS_2$). With the expected monotonic increase of the exciton peak intensity with T up to 150 K, there is an overall reduction of the peak intensity as $l$ increases, with a crossover around 150 K, afterward, it reverses for $WS_2$. The full spectrum of the PL intensity variation for $l = 0; 4$ at 150 K and 200 K is shown to demonstrate the reversal behaviour in Figure 3b. To gain a comprehensive understanding of T-dependent intensity variation, we consider a rate equation model[36,37] involving the bright $X_K^b$ ($X_{K'}^b$) and dark excitons $X_K^d$ ($X_{K'}^d$) with their populations $N_b$ ($N_b'$), and $N_d$ ($N_d'$) in $K$ ($K'$) valley, as depicted in Figure 3c. We assume that $X_K^b$ ($X_{K'}^b$), and $X_K^d$ ($X_{K'}^d$) are generated with rates of $g_K^b$ ($g_{K'}^b$) and $g_K^d$ ($g_{K'}^d$) respectively via steady-state excitation with radiative and the non-radiative recombination time of $\tau_b$ and $\tau_d$ respectively. The transition, $X^b \rightarrow X^d$ ($X^d \rightarrow X^b$) occurs due to intravalley spin-flip scattering with a rate $\frac{1}{\tau_{bd}}\left(\frac{1}{\tau_{bd}}e^{\frac{-\Delta E}{k_B T}}\right)$, where $\Delta E$ is the energy difference between $X^b$ and $X^d$ states.

The rate equation capturing the scattering mechanism of our interest can be written as follows: Detailed calculation is shown in Supporting Information.

$$\frac{d}{dt}\binom{N}{N'} = \begin{bmatrix} A & B \\ B & A \end{bmatrix}\begin{bmatrix} N \\ N' \end{bmatrix} + \begin{bmatrix} G \\ G' \end{bmatrix}$$

Where, $N = \binom{N_b}{N_d}$, $N' = \binom{N'_b}{N'_d}$, $G = \binom{g_K^b}{g_K^d}$ and $G' = \binom{g_{K'}^b}{g_{K'}^d}$ are populations and the generation rates as defined above. $A$ is a $2 \times 2$ matrix representing the intravalley scattering describing the interaction between $X^b$ and $X^d$ states within the valley. The $2 \times 2$ matrix, $B$ represents the intervalley scattering rate, describing the interaction between two valleys. Since we have used linearly polarized light, hence both the degenerate valleys are excited equally and contribute equally to the PL emission. Therefore, intervalley scattering process does not affect the intensity variation. Thus, any contribution from intervalley dark-excitons [11,38] may not alter the interpretation of PL intensity variation that we observe with $l$. Consequently, the effect of matrix $B$ only becomes important to understand the valley depolarization and decoherence effect, which we have investigated in detail in monolayer WSe$_2$ [39]. The solution of bright exciton intensity can be written as,

$$I_{bright} = \frac{2\left(e^{\frac{\Delta E}{K_B T}}g_b\tau_{bd}+(g_b+g_d)\tau_d\right)}{e^{\frac{\Delta E}{K_B T}}(\tau_b+\tau_{bd})+\tau_d} \quad (1)$$

From our analysis, we observe that at low T, $\left(\frac{N_b}{N_d}\right) > e^{\frac{-\Delta E}{k_B T}}$, implying excitons are not thermalized. However, the thermalization process becomes faster due to an increase in the intravalley scattering rate as $q$ increases, allowing more $X^b \rightarrow X^d$ transitions to bring $\frac{N_b}{N_d}$ closer to the Boltzmann ratio, resulting in the intensity decreasing with $l$ in low T regime. On the other hand, $|\Delta g_d(l)| > |\Delta g_b(l)|$ with increasing $l$ as mentioned above further drives the system closer to equilibrium at low T.

At a high T regime, the system thermalizes within the PL time scale ($\tau_b$), i.e., $\frac{N_b}{N_d}$ follows the Boltzmann ratio. Therefore, the effect of increasing intravalley scattering rate does not affect the system. However, since $|\Delta g_d(l)| > |\Delta g_b(l)|$, excess of $X^d$ is converted into $X^b$ to maintain the Boltzmann ratio at high-T regime, leading to the increasing PL intensity with $l$. The crossover region is estimated at around 120 K using our rate equation model as shown in the vertical offset in Figure 3a. However, experimentally we find this crossover region between 150 K to 200 K. One can improve the estimation of the crossover region by incorporating EHEI in a modified energy dispersion of the bright excitons[28] in the rate equation model. A similar trend in the PL intensity variation is observed in WSe$_2$, as shown Figure S9 in Supporting Information. Around 250 K, the PL intensity variation starts reversing with $l$, but for higher $l$ the variation is less prominent. It is expected that in case of WSe$_2$ the increasing trend of PL intensity variation may become distinct much above the room temperature value. Room T PL intensity variation for WSe$_2$ is presented in Supporting Information Figure S10. To elucidate the dominating behaviour of intravalley scattering and field variation in low and high-T regimes, we have considered their individual effect separately as described below.

If only the impact of intravalley scattering rate variation is considered, by keeping generation rate ($g_b$ and $g_d$) constant with $l$ in equation (1), then the fitted curve has a negligible change at a high T regime, whereas, at a low T regime, PL intensity is reduced noticeably due to decrease in $\tau_{bd}$ with $l$ as shown in Figure 3d. On the other hand, if only the field variation is considered, then PL intensity increases noticeably in high T regime. There is also reduction in PL at low T regime, but the intensity variation is one order less than empirical data with crossover around 60 K as shown in vertical offset in Figure 3d. At low T, $\left(\frac{N_b}{N_d}\right) > e^{\frac{-\Delta E}{k_B T}}$, implying a decrease in $N_b$ further drives the system towards equilibrium. Therefore, the crossover of PL intensity variation can be understood as an effect of modification of of electric field intensity and intravalley scattering rate due to momentum transfer into excitonic systems.

**Conclusion**

We utilize tunable dispersion of optical vortex beam to transfer momentum from twisted light to exciton, which alters the intravalley spin-flip scattering processes responsible for the PL intensity variation in W-based TMDs. At a fixed temperature, contributions of spin-flip scattering due to exciton-phonon and impurity scattering are treated as a constant background without affecting our results on exciton-momentum dependent relaxation processes, tunable by topological charge. We demonstrated that the PL intensity can be varied by changing excitons momentum without changing bath temperature and excitation power. Such interface of vortex beam optics with the centre of mass motion of excitons, influencing dynamics of internal degrees of freedom in 2D materials can be a promising technique to explore novel phenomena in exciton transport and spin-valley physics.

**Methods**

Monolayers of $WSe_2$ and $WS_2$ were mechanically exfoliated from commercially available bulk crystals and dry transferred on to a $SiO_2$/Si substrate. Optical Vortex beam was generated using an amplitude only spatial light modulator utilizing computer-generated holograms. Measurements were performed using a closed cycle optical microscopy cryostat (Montana Instruments) with variable temperature range of 3.2 K to 295 K. Princeton Instruments spectrometer (SP2750) and a liquid nitrogen cooled detector (PyLoN:400BR-eXcelon) was used to measure the PL signal.

**Acknowledgments**

We acknowledge Sourin Das, Mandar M. Deshmukh, and Chitraleema Chakraborty for their valuable comments on this work.

**Funding Sources**

S. D. acknowledges Science and Engineering Research Board (SB/S2/RJN-110/2016, CRG/2018/002845), STARS (MoE/STARS-1/647), Indian Institute of Technology Kharagpur (IIT/SRIC/ISIRD/2017-2018), and Ministry of Human Resource Development (IIT/SRIC/PHY/NTS/2017-18/75) for the funding and support for this work. D. C. acknowledges Council of Scientific and Industrial Research, JRF (09/081(1352)/2019-EMR-I) for the financial assistance.

**ASSOCIATED CONTENT**

**Supporting Information**

The Supporting Information is available.

Details of experimental setup for PL, PL intensity variation with $l$ at 50 K, Sample homogeneity, Excitation and collection efficiency, Reproducibility in other $WSe_2$ monolayer samples, PL intensity variation with $l$ and $T$ for $WSe_2$, Room temperature PL intensity variation for WSe2 with $l$, Rate equation model for temperature dependence PL intensity for OV beam, and Electric field distribution at the focal plane for OV beam excitation.

**Figure 1**

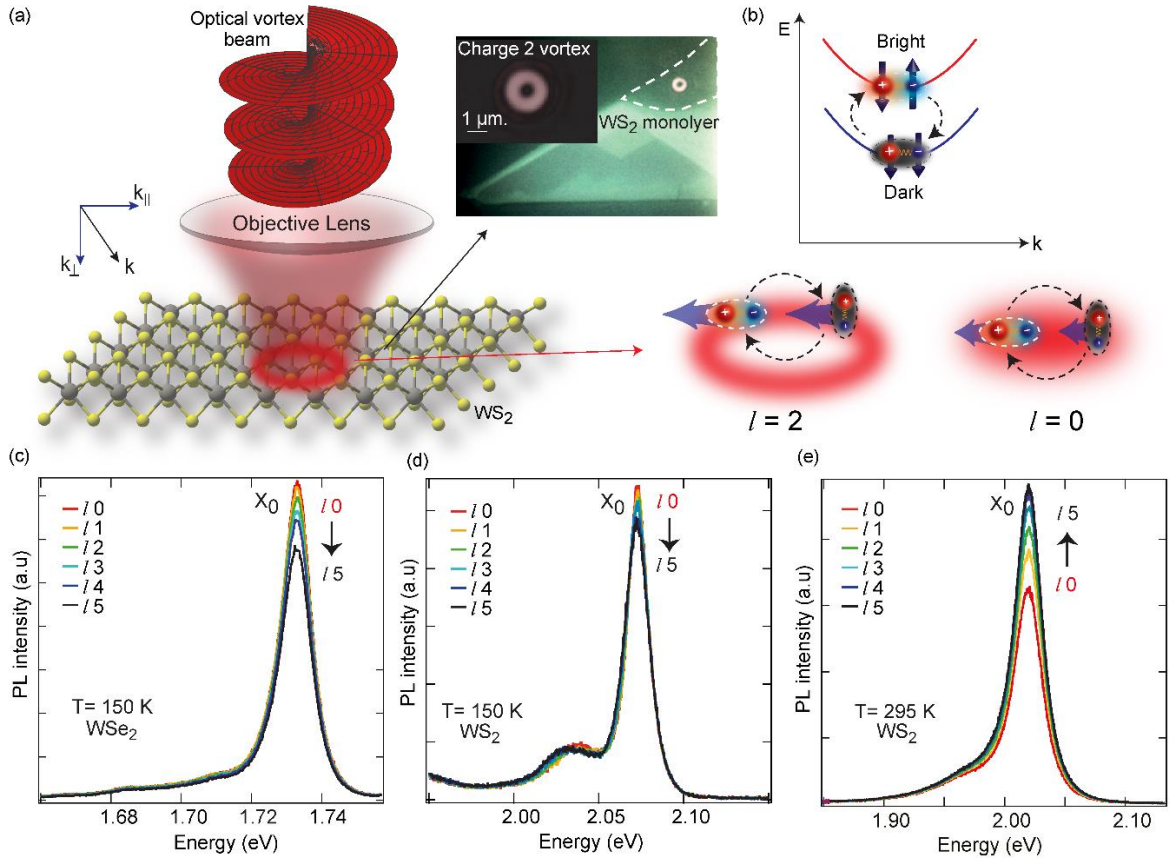

**Figure 1.** (a) Schematic of focused OV beams incident on monolayer $WS_2$ sample, with zoomed view optical image for $l = 2$ as the inset. (b) The momentum of excitons (represented with blue arrows) increase with topological charge of excitation beam as represented by higher magnitude blue arrows for $l = 2$ than $l = 0$, facilitating the increased intravalley scattering as shown in band diagram representation. (c) and (d) PL spectra of $WSe_2$ and $WS_2$ monolayer for different $l$ at T = 150 K. PL intensity decreases with increasing $l$ (indicated by the black arrow). (e) PL spectrum for $WS_2$ monolayer at 295 K, showing reversal of intensity variation which is increasing as $l$ increases (indicated by the black arrow).

**Figure 2**

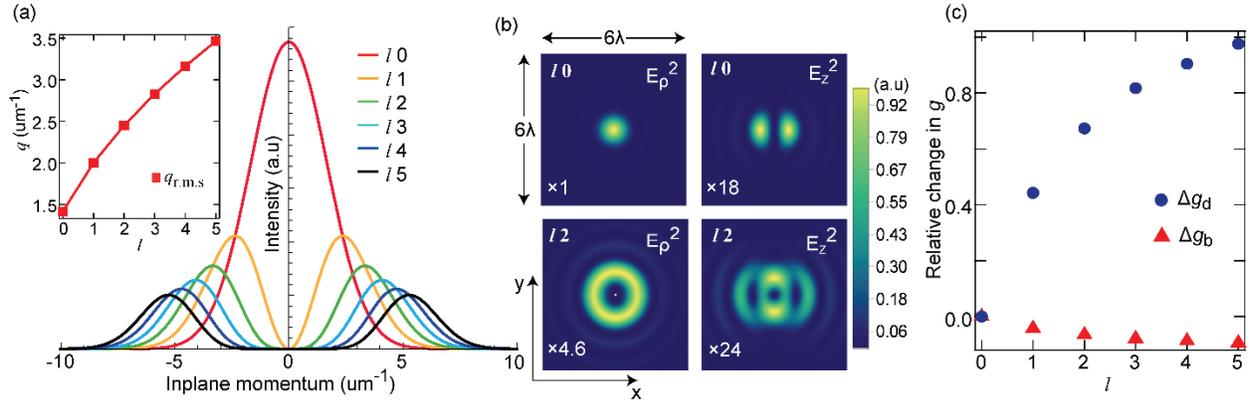

**Figure 2.** (a) The simulated intensity distribution of OV beam for different $l$ in $k_{\parallel}$ space on the focal plane by angular spectrum representation with variation of excitons $q_{r.m.s}$ with $l$. (b) Real space intensity distribution of simulated $|E_z|^2$ and $|E_\rho|^2$ at focal plane for $l$ 0 and $l$ 2. All figures have the same x and y axes ranging from $-3\lambda$ to $+3\lambda$ (c) The relative change in generation rate of bright and dark excitons ($\Delta g_b, \Delta g_d$) for OV beams excitation in comparison to Gaussian beam.

**Figure 3**

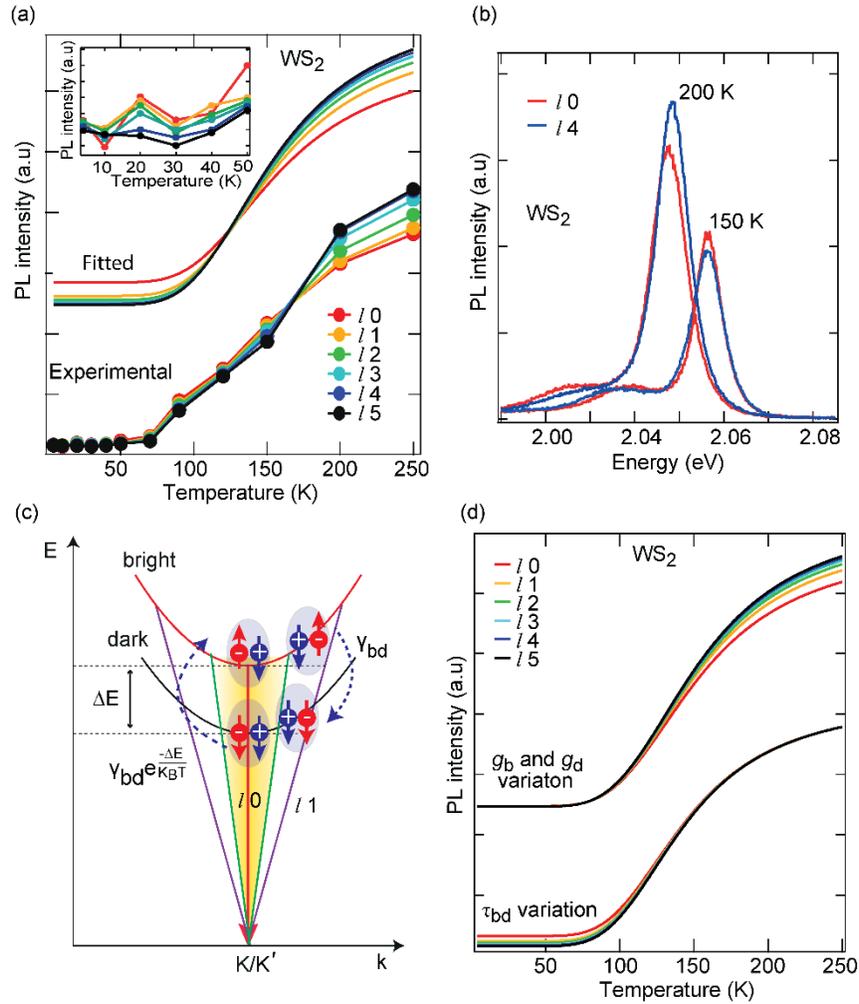

**Figure 3.** (a) Exciton intensity variation with T for WS$_2$ monolayer showing crossover regime between 150 K and 200 K. PL intensity decreases up to 150 K and afterwards starts increasing with increasing $l$. Fitted exciton intensity variation with T for different $l$ is plotted with a vertical offset, showing crossover regime around 125 K. Inset is showing intensity variation up to 50 K (b) PL spectra for $l$ = 0 and 4 showing reversal in trend at 150 K and 200 K. (c) Schematic diagram for the rate equation model presenting bright and dark exciton states in solid red and black curves. Curved arrows indicate the intravalley spin-flip scattering processes considered in the model. (d) Simulated plots for comparison between the individual effect of the $\tau_{bd}$ and ($g_b$, $g_d$) variation (shown in vertical offset) on T dependent PL intensity variaion.

# Supporting Information

# Probing spin dynamics of 2D excitons with twisted light

Aswini Kumar Pattanayak[†], Pritam Das [†], Devarshi Chakrabarty, Avijit Dhara, Shreya Paul, Satyait Maji, Maruthi Manoj Brundavanam , Sajal Dhara[*]

*Department of Physics, Indian Institute of Technology, Kharagpur-721302, India*

**Experimental setup for PL:**

The optical vortex (OV) beam of the desired $l$ is prepared by shining a Gaussian beam on the SLM, which displays the corresponding computer-generated hologram (CGH)[1]. The CGHs for generating $l$ from 0 to 2 have been shown in Fig. S1. The 1st order diffracted beam, which contains the OV beam of desired $l$, is focused by a microscope objective (MO) on the monolayer sample. For PL measurement, the excitation beam is made plane-polarized by passing it through Pol 1, as shown in Figure S1. The emitted PL signal is collected through the same MO and is directed into the spectrometer.

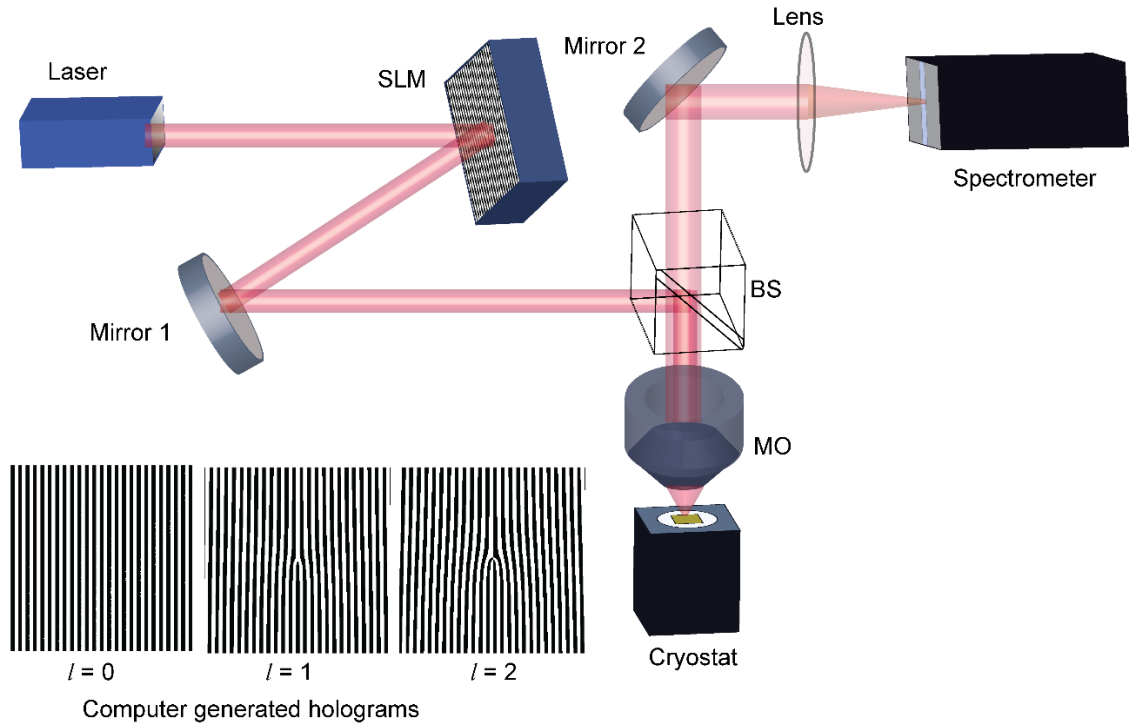

**Figure S1**: **Experimental setup PL measurement using OV beam:** Display of amplitude only SLM with computer-generated holograms (CGHs) used to generate optical vortices are shown. The sample is placed inside the cryostat, excited through a microscope objective (MO) lens (0.7 NA). The output PL emission is collected through the same MO and is focussed by a lens into a spectrometer slit.

**PL intensity variation with $l$ at 50 K**

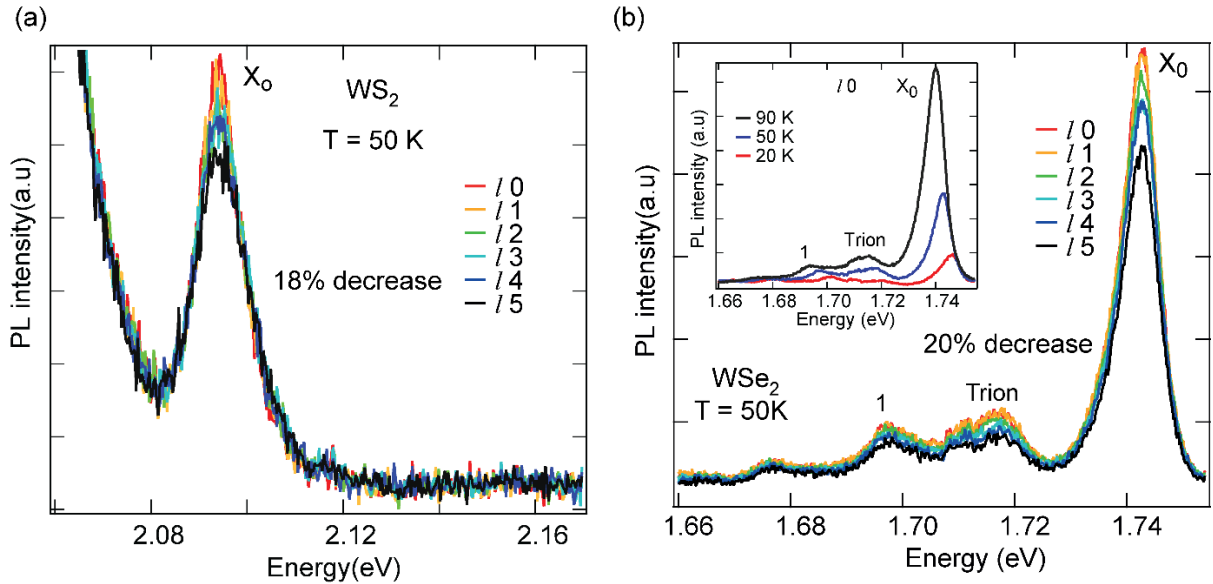

**Figure S2**: **PL spectrum for WS$_2$ and WSe$_2$ monolayer for OV beam at 50 K:** (a,b) The peak around 2.098 eV and 1.742 eV is identified as bright exciton (X$_0$) for WS$_2$ and WSe$_2$ monolayer respectively. For WSe$_2$ monolayer, the peak appearing at 1.710 eV has been assigned as trion peak as reported in literature [2]. And the peak at 1.697 eV can be a complex bright exciton species as its line width is larger than X$_0$ and the peak intensity also increasing with temperature as shown in the inset of Figure S2(b).

**Sample Homogeneity:**

We have performed same experiments to verify the effect of inhomogeneity using freshly exfoliated samples at low and high-temperature regimes, as shown in the figures below. From our measurements, we have observed that the effect of intensity variation with $l$ is dominant over the intensity fluctuations arising from inhomogeneity in the concerned region of the sample.

It can be observed that the diameter of the vortex beam changes from ~1.5 $\mu m$ to ~3 $\mu m$ when the topological charge ($l$) varies from 0 to 4. To verify the inhomogeneity around this region we have measured the PL intensity using a Gaussian beam at 4 different points within the circle of diameter ~3 $\mu m$ as shown in white color on samples (grid dimension ~ 2 $\mu m$ x 2 $\mu m$) in Figure S3(a) below. We indeed found the PL intensity variation due to sample inhomogeneity to be around 7% (Figure S3(b)) with respect to the central position (Pos 0), whereas we have observed 25 % increase (Figure S3(c)) in PL intensity with a change in $l$ from 0 to 4 at 295 $K$.

At 120K, the PL intensity variation due to inhomogeneity is negligible, which makes the sample almost homogeneous around the region of interest, as shown in Figure S3(e). A decrease in PL intensity of ~12% (Figure S3(f)) is observed for change in $l$ from 0 to 4, which demonstrates the reproducibility of our results. Similarly, we have also recorded other positions and have observed the effect of PL intensity variation with $l$ dominant over the spatial inhomogeneity.

To further rule out inhomogeneity as a dominant factor, we have performed PL intensity variation with $l$ on the bilayer region presented in sample 2 as shown in Figure S3(d). We have not observed noticeable intensity variation for different OAM of OV beams. Therefore, it can be concluded that the PL intensity variation on the monolayer region comes primarily from the variation of exciton dynamics caused by the OV beam.

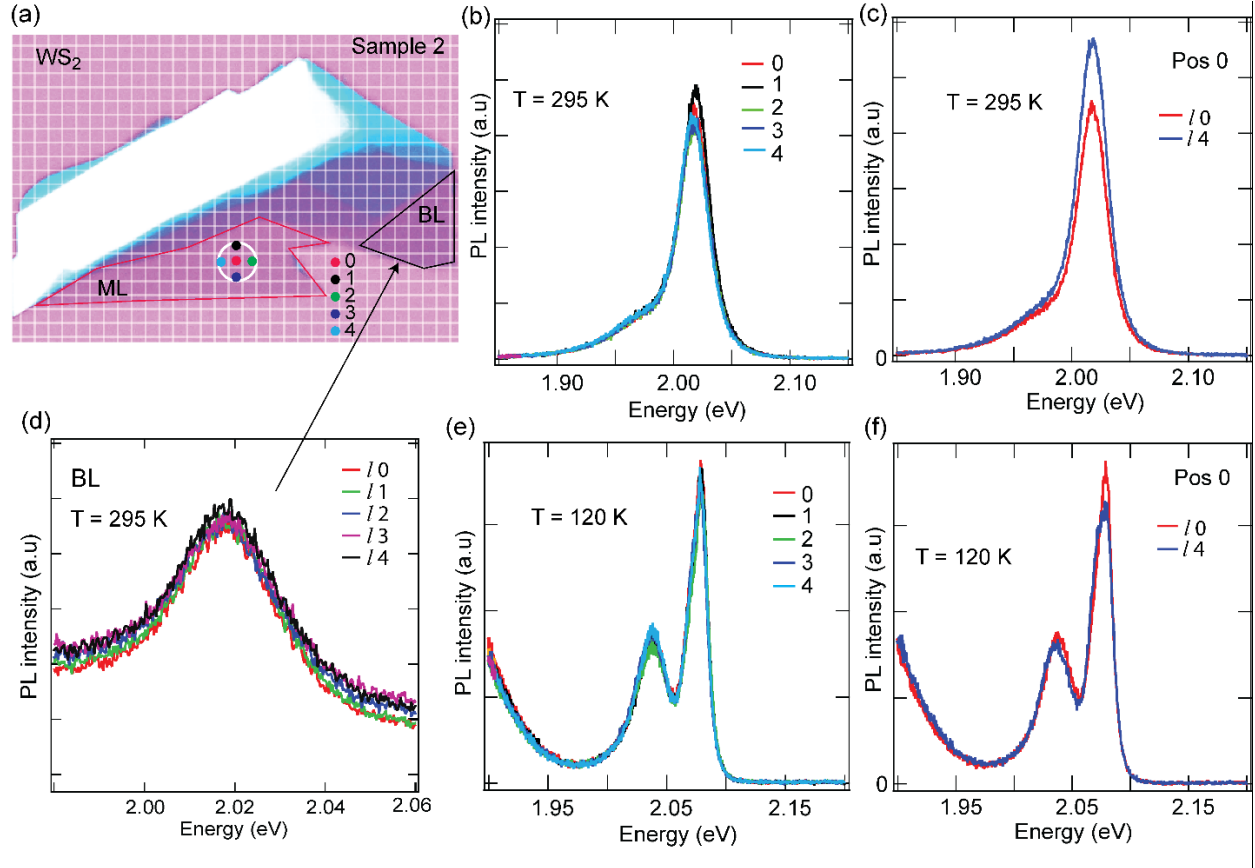

**Figure S3:** (a) Gridded optical image of $WS_2$ sample 2. (b,c) Comparison between sample inhomogeneity and $l$ variation in PL spectrum at 295 K. (d) PL intensity variation with l on bilayer $WS_2$. (e,f) Comparison between sample inhomogeneity and $l$ variation in PL spectrum at 120K.

We have also performed the same experiment on a different sample (3) (grid dimension ~ 2 $\mu m$ x 2 $\mu m$) at 295 K as shown in Figure S4(a). The sample is almost homogeneous in the concerned region (white circle with 3 $\mu m$ diameters), as shown in Figure S4(b), whereas we have obtained 20% increase (Figure S4(c)) in PL intensity by changing $l$ from 0 to 4.

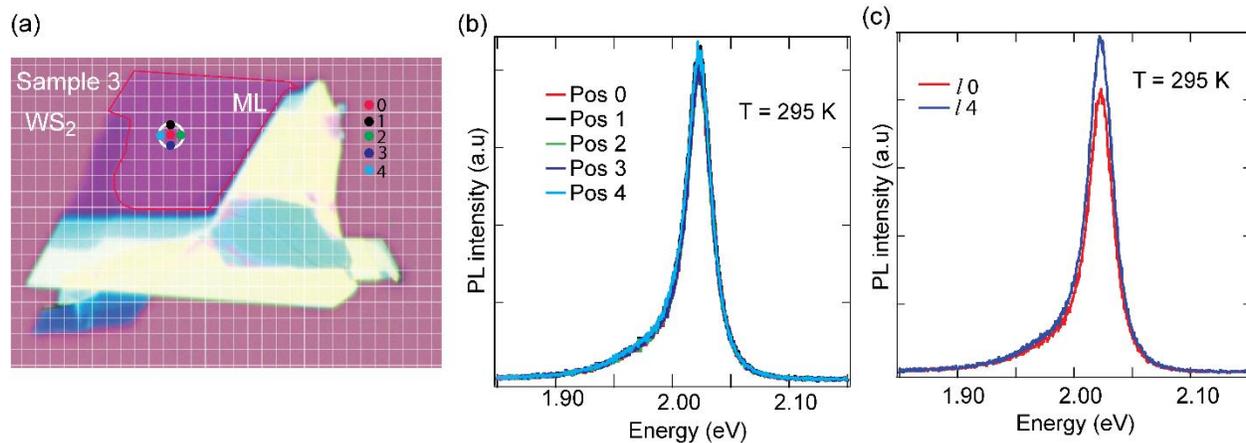

**Figure S4**: (a) Gridded optical image of WS$_2$ Sample 3. (b,c) Comparison between sample inhomogeneity and $l$ variation in PL spectrum at 295 K

The experimental observations shown in the manuscript have been performed on sample 1 as shown in Figure S5(a). For instance, the homogeneity was performed only at 30 K a very close vicinity. It can be well observed the spatial exciton peak intensity variation remains almost constant, as shown in Figure S5(b).

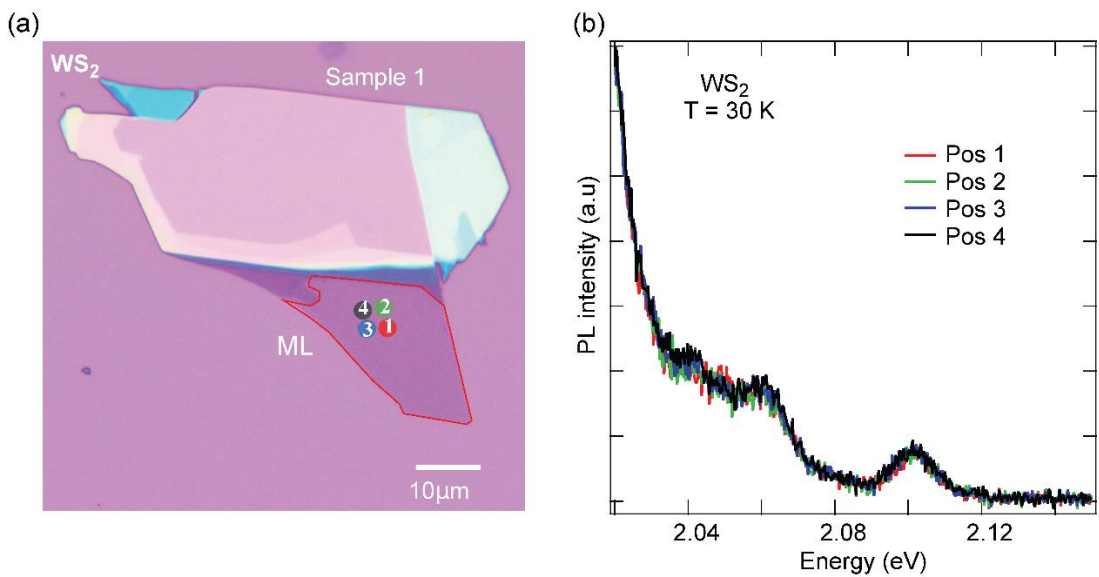

**Figure S5:** (a) Optical image of WS2 sample 1. (b) PL spectrum at different positions showing sample homogeneity at 30 K.

Additionally, the PL image for different OAM beams obtained from the CCD camera has been shown in Figure S6. We find the spatial variation of PL emission of the sample for different OAM beams is largely uniform.

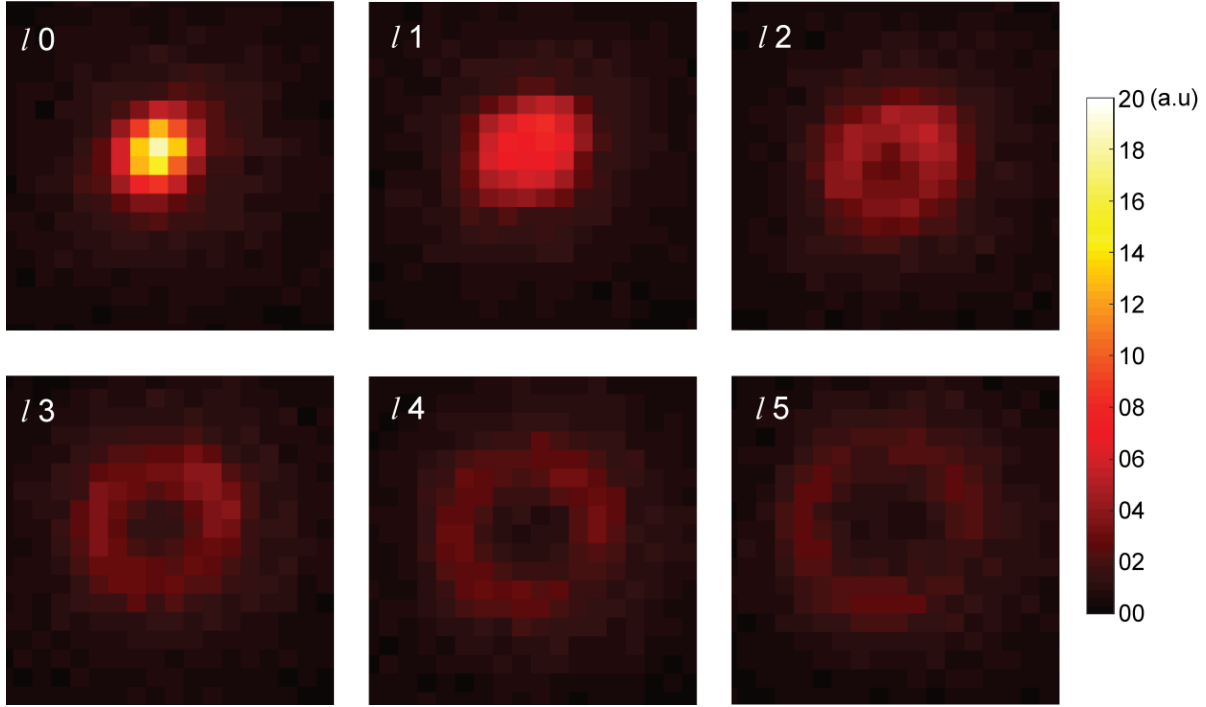

**Figure S6**: Spatial PL image recorded from CCD camera for $l = 0$ to 5 on monolayer $WS_2$.

**Excitation and collection Efficiency :**

For calibration propose, we have measured the reflection of OAM beam from $SiO_2$ substrate as presented in Figure S7(a) with normalizing for incident power. The detected reflected light intensity shows nearly no change with $l$ number variation (Figure S7(b)), suggesting equal excitation and collection efficiency for different OAM beams up to $l = 5$.

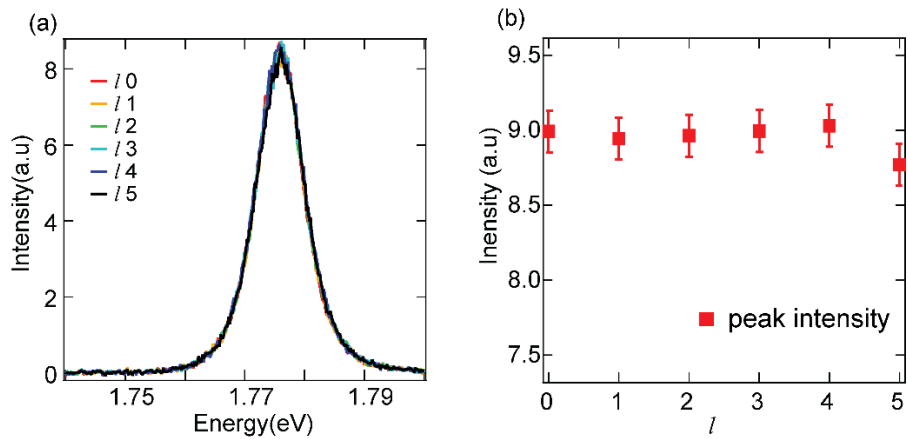

**Figure S7:** (a,b) Collection efficiency of different order OV beam after getting reflected from $SiO_2$ substrate.

## Reproducibility in other WSe$_2$ monolayer samples

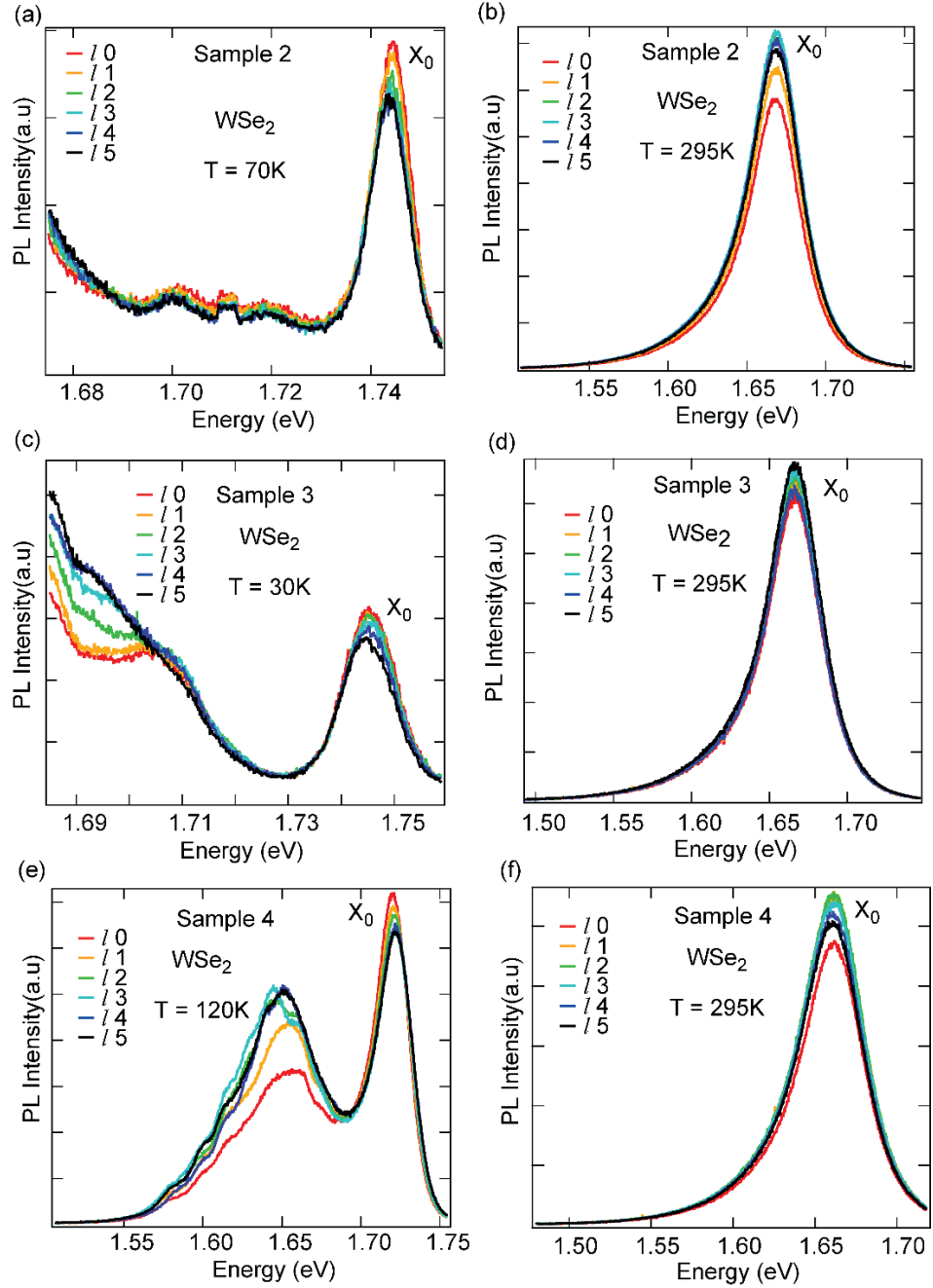

**Figure S8: Different WSe$_2$ monolayer samples PL spectrum for OV beams at low and room temperature:** (a,c,e) Exciton peak (X$_0$) is decreasing with increasing $l$ at low-temperature regime for sample 2- 4. (d,e,f) At room temperature the variation of X$_0$ is less prominent with change in $l$ for samples 2-4.

## PL intensity variation with *l* and T for WSe$_2$

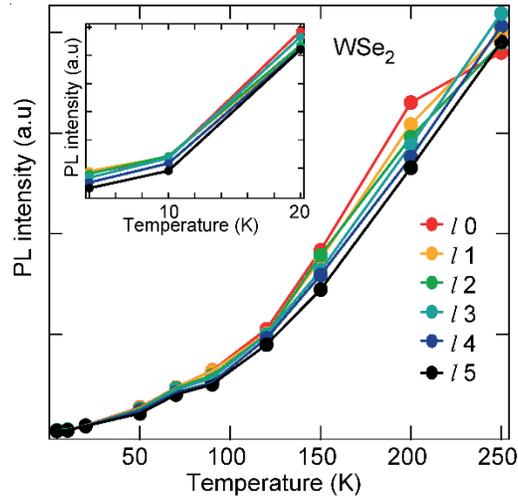

**Figure S9:** Variation of exciton intensity with T showing the reduction of PL intensity with increasing *l* for WSe$_2$ monolayer. Inset shows PL intensity variation up to 20 K.

## Room temperature PL intensity variation for WSe$_2$ with *l*

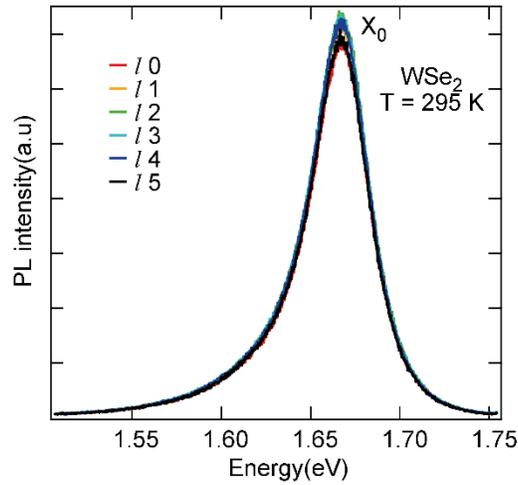

**Figure S10**: PL spectra for WSe$_2$ at room temperature (295 K) for OV beams of different *l*. The increasing behavior of exciton for higher-order *l* number is not so pronounced.

**Rate equation model for temperature dependence PL intensity for OV beam:**

The following rate equations [3,4] are used to model the intensity variation with temperature for different $l$. Schematic diagram for rate equation model presenting bright and dark exciton states with different valleys in the W-based TMDs system is shown in Figure 2c in the manuscript. The rate equation model for the W-based TMDs system can be written as,

$$\frac{d}{dt}\begin{pmatrix} N_b \\ N_d \\ N_b' \\ N_d' \end{pmatrix} = \begin{pmatrix} -\left(\frac{1}{\tau_b}+\frac{1}{\tau_{bd}}+\frac{1}{\tau_{KK'}}\right) & \frac{1}{\tau_{bd}}e^{\frac{-\Delta E}{K_B T}} & \frac{1}{\tau_{KK'}} & 0 \\ \frac{1}{\tau_{bd}} & -\left(\frac{1}{\tau_d}+\frac{1}{\tau_{bd}}e^{\frac{-\Delta E}{K_B T}}\right) & 0 & 0 \\ \frac{1}{\tau_{KK'}} & 0 & -\left(\frac{1}{\tau_b}+\frac{1}{\tau_{bd}}+\frac{1}{\tau_{KK'}}\right) & \frac{1}{\tau_{bd}}e^{\frac{-\Delta E}{K_B T}} \\ 0 & 0 & \frac{1}{\tau_{bd}} & -\left(\frac{1}{\tau_d}+\frac{1}{\tau_{bd}}e^{\frac{-\Delta E}{K_B T}}\right) \end{pmatrix}\begin{pmatrix} N_b \\ N_d \\ N_b' \\ N_d' \end{pmatrix} + \begin{pmatrix} g_K^b \\ g_K^d \\ g_{K'}^b \\ g_{K'}^d \end{pmatrix}$$

Whereas $N_b$, $N_d$ are the bright and dark exciton population in $K$ valley and $N_b'$, $N_d'$ are the dark exciton population number in $K'$ valley. $\tau_b$, $\tau_d$ are the radiative and non-radiative recombination time for bright and dark excitons, respectively. $g_K^b$, $g_K^d$ are the bright and dark exciton generation rate in $K$ valley and $g_{K'}^b$, $g_{K'}^d$ are the bright and dark exciton generation rate in $K'$ valley. $\frac{1}{\tau_{bd}} = \gamma_{bd}$ and $\frac{1}{\tau_{bd}}e^{\frac{-\Delta E}{k_B T}} = \gamma_{bd}e^{\frac{-\Delta E}{k_B T}}$ are the scattering rate from bright state to dark state transition and dark state to bright state transition in W-based TMDs system, respectively. $\frac{1}{\tau_{KK'}} = \gamma_{KK'}$ is the intervalley scattering rate between two valleys.

The above 4 by 4 matrix can be written as,

$$\frac{d}{dt}\begin{pmatrix} N \\ N' \end{pmatrix} = \begin{bmatrix} A & B \\ B & A \end{bmatrix}\begin{bmatrix} N \\ N' \end{bmatrix} + \begin{bmatrix} G \\ G' \end{bmatrix}$$

Where A and B are $2 \times 2$ matrices representing the intravalley and intervalley scattering process, respectively. A and B can be written as

$$A = \begin{bmatrix} -\left(\frac{1}{\tau_b}+\frac{1}{\tau_{bd}}+\frac{1}{\tau_{KK'}}\right) & \frac{1}{\tau_{bd}}e^{\frac{-\Delta E}{K_B T}} \\ \frac{1}{\tau_{bd}} & -\left(\frac{1}{\tau_d}+\frac{1}{\tau_{bd}}e^{\frac{-\Delta E}{K_B T}}\right) \end{bmatrix} \text{ and } B = \begin{bmatrix} \frac{1}{\tau_{KK'}} & 0 \\ 0 & 0 \end{bmatrix}$$

For CW excitation at equilibrium,

$$\frac{dN_b}{dt} = 0, \frac{dN_d}{dt} = 0, \frac{dN_b'}{dt} = 0, \frac{dN_d'}{dt} = 0$$

For the PL measurements as linearly, polarized light has been used so,

$g_K^b = g_{K'}^b = g_b$ and $g_K^d = g_{K'}^d = g_d$ due to equal excitation in both valleys.

The solution of bright exciton intensity can be written as,

$$I_{bright} = I_K^b + I_{K'}^b = \frac{2\left(e^{\frac{\Delta E}{K_B T}} g_b \tau_{bd} + (g_b + g_d)\tau_d\right)}{e^{\frac{\Delta E}{K_B T}}(\tau_b + \tau_{bd}) + \tau_d} \quad (1)$$

For near-resonant excitation, the cascade-like exciton relaxation process due to phonon scattering will cause a mixture of coherent and non-coherent excitons, as evidenced by valley coherence and polarization measurements. In the WSe2 sample, we have observed 56% and 24% valley coherence and polarization [5], respectively, for a 660 nm laser excitation (130meV above the A exciton resonance). Similarly, 16% valley polarization has been reported in WS$_2$ sample for 532nm laser excitation (285meV above the A exciton resonance)[6,7]. This indicates that excitons are not fully relaxed due to phonons scattering and reserve the phase information about the polarization state of the excitation beam. Consequently, the coherent excitons will follow the photons' momentum distribution, which has been used in the rate equation model to explain the PL intensity variation with $l$ of the vortex beam.

In our experiments, the exciton-phonon relaxation cascade due to excess energy acts as constant background in the PL intensity variation with $l$ number. For near-resonant excitation, the generation rate of bright exciton in the rate equation model can be expressed as $g_b = \beta e^{\frac{-\Delta \epsilon}{K_B T}} g_{b'}$. Where, $\Delta \epsilon$ is the energy detuning or excess energy, and $g_b$, $g_{b'}$ are the exciton generation rate at the bottom and higher energy point in the bright exciton band respectively. Thus, in our model the excess energy can be incorporated by treating $g_{b'}$ as a fitting parameter. Our observation of finite degree of valley coherence and polarization at near resonant excitation (which is $\Delta \epsilon$ above the minimum of excitonic band) suggests that the excess energy is only responsible for the partial depolarization and decoherence effect which also reduces the effect of variation of PL intensity with the optical vortex charge.

*Fitting of intensity variation data using equation (1):*

The number of parameters used in the rate equation model has been minimized while considering all physically important processes. The energy difference between bright and dark exciton state $\Delta E$ is fixed to 55 meV for WS$_2$ monolayer[8]. From the five parameters $(g_b, g_d, \tau_{bd}, \tau_d, \tau_b)$ used in the above mode, $\tau_b$ and $\tau_d$ has been kept fixed for different $l$.

The dependence of $\tau_{bd}$ $\left(\frac{1}{\gamma_{bd}}\right)$ on $q$ can be calculated by Andrada e Silva and La Rocca mechanism [9]. In the mechanism, the Hamiltonian with spin basis [Bright exciton $\{(\uparrow_e, \downarrow_h), (\downarrow_e, \uparrow_h)\}$, Dark Exciton $\{(\downarrow_e, \downarrow_h), (\uparrow_e, \uparrow_h)\}$] has an off-diagonal matrix element between Bright and Dark exciton states. These off-diagonal terms are responsible for the coupling between Bright and dark exciton spin state with the finite center of mass momentum ($q$). Which can be represented as an effective pseudo magnetic field in the x-y plane. By the interaction between these pseudo magnetic field with the electron's spin, the spin-flip is happening. Which can be given as[9]

$$\gamma_{bd} = \frac{4\alpha_{so}^2}{\hbar} \frac{\tau}{1+\left(\frac{\Delta_0 \tau}{\hbar}\right)^2} q^2$$

So, we use $\gamma_{bd} \propto q^2$ in the rate equation model.

Intravalley spin-flip scattering time $\tau_{bd}$ can be written as $\tau_{bd} \propto \frac{1}{(q)^2}$.

$g_b$ and $g_d$ are proportional to $|E_\rho|^2$ and $|E_z|^2$ respectively as mentioned in manuscript, which can be presented as $g_b = C_1|E_\rho|^2$ and $g_d = C_2|E_z|^2$, where $C_1$ and $C_2$ are proportionality constants.

Equation (1) has been simultaneously fitted with the experimental data of temperature-dependent PL intensity variation for WS$_2$, as shown in figure 2a in the manuscript. The fitted curves are shown in figure 2d in the manuscript.

The fitted parameters are,
$\tau_b = 10\ ps,\ \tau_d = 539.2\ ps$

**Table S1:** Fitted parameters for WS$_2$ PL intensity variation.

| $l$ | $g_b$ | $g_d$ | $\tau_{bd}$ (ps) |
|---|---|---|---|
| l0 | 8.19 | 2.97 | 1.57 |
| l1 | 7.83 | 4.29 | 0.77 |
| l2 | 7.65 | 4.97 | 0.53 |
| l3 | 7.54 | 5.40 | 0.39 |
| l4 | 7.47 | 5.66 | 0.32 |
| l5 | 7.41 | 5.87 | 0.26 |

From Table S1, the fitted parameters $g_b$ and $g_d$ only represent mathematical fitting quantity and should not be scaled to experimental parameters as the PL intensity is considered in arbitrary unit (a.u). Here we would like to clarify that the generation rates obtained from PL intensities should not be compared with the oscillator strengths of excitons. The theoretically predicted ratio of oscillator strength of bright to dark exciton is nearly 1000 [10], which can be estimated from reflection or absorption studies. However, oscillator strength cannot be determined directly from PL spectra due to the unknown population of $X_0$ and $X_d$ at near or off-resonant conditions. The estimated intravalley scattering time (as presented in Table S1 in supporting information) is of the same order of magnitude as reported in the literature for WS$_2$ monolayer [11].

From Table S1, it can be observed that $\tau_{bd}$ is decreasing, and the sum of $g_b$ and $g_d$ is increasing with increasing $l$. Decreasing $\tau_{bd}$ is responsible for reduction PL intensity at low temperatures and increasing $(g_b+g_d)$ is responsible for the enhancement of the PL intensity at the high-temperature regime. The physical explanation has been given in the manuscript. Our model explains that variation of Intravalley scattering rate and generation rate are two main phenomena responsible for the unique PL intensity variation with crossover regimes for different $l$. If one of the above phenomena is not considered, the fitted curve highly deviates from the experimental results.

In our model, We have not considered the T- dependent radiative and non-radiative lifetime since the exact functional form for whole temperature range (4-300K) for $\tau_b$ and $\tau_d$ is not known to our knowledge. Although at high-temperature regime, radiative lifetime can be written as, $\tau_{rad}^{eff} = \frac{3}{2}\frac{K_B T}{E_0}\tau_{rad}^0$ [12], this is valid when $K_B T$ is the order of bright and dark exciton splitting (Arise from the thermalized exciton). Still if we consider linear T-dependence of radiative and non-radiative life time for whole temperature range i.e $\tau_b(T) = \tau_{b0}T$ and $\tau_d(T) = \tau_{d0}T$ in our model, then the crossover of the fitted curve arises around 85 K as shown below.

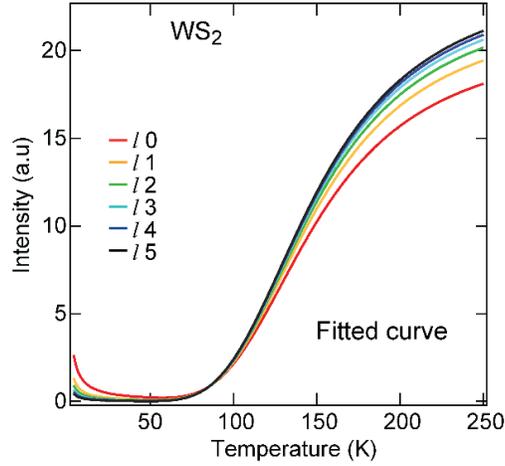

**Figure S11**: Fitted intensity variation with temperature with different l for WS2 sample by considering linear T dependence of $\tau_b$ and $\tau_d$.

**Electric field distribution at the focal plane for OV beam excitation:**

The Electric field distribution at the focal plane which has been used to calculate the integrated $|E_\rho|^2 = |E_x|^2 + |E_y|^2$ and $|E_z|^2$ as shown below for $l = 0$ to $l = 2$ [13].

Effective $N.A = 0.7$, focal length $= 1.53$ mm, wavelength $(\lambda) = 520$ nm.

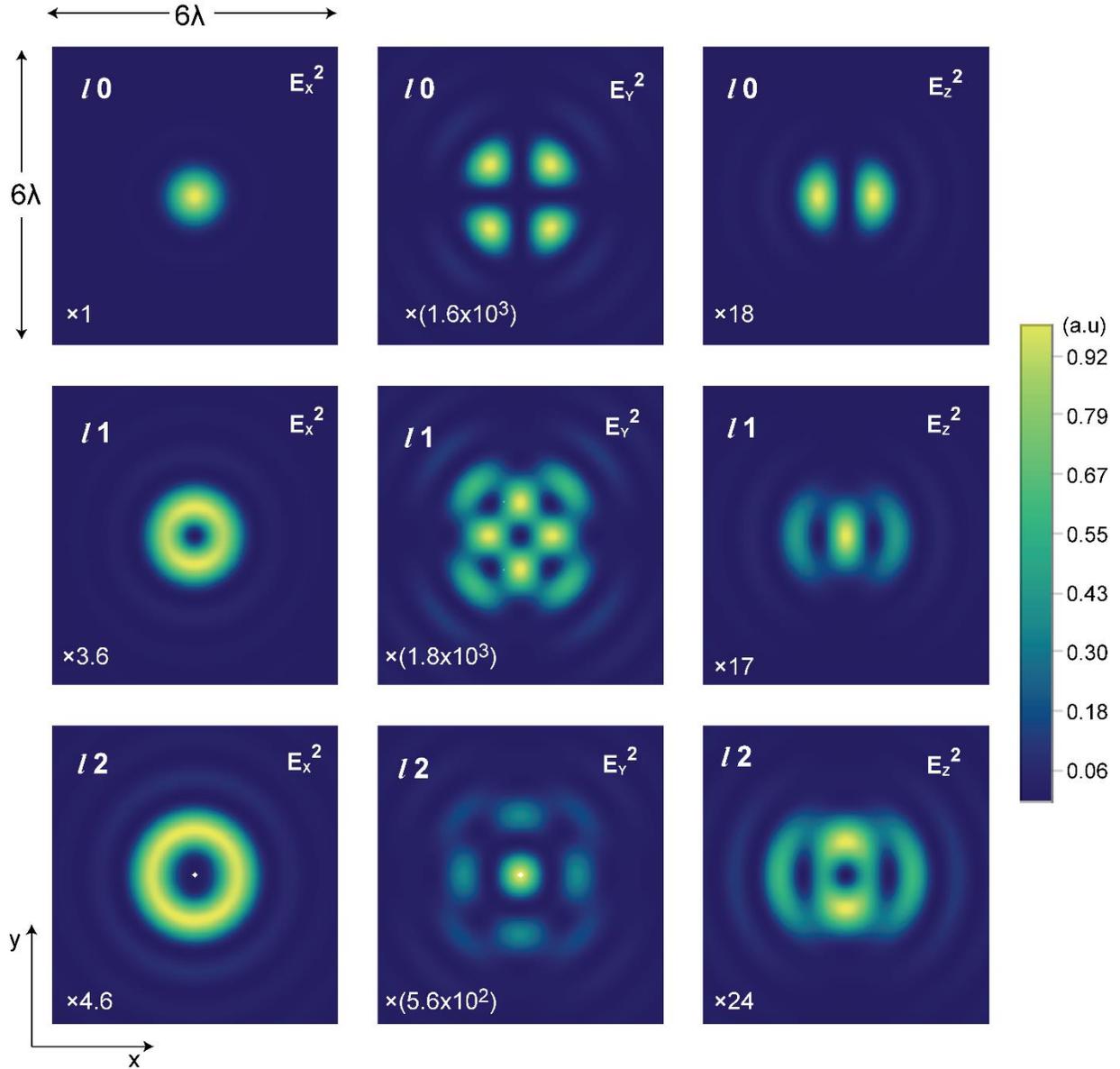

**Figure S12:** Electric field distribution for $|E_x|^2, |E_y|^2, |E_z|^2$ for $l = 0$ to 2 with their normalization factor. All figures have the same x and y axes ranging from $-3\,\lambda$ to $+3\,\lambda$.